\documentclass[aps,preprint,tightenlines,showpacs,nofootinbib]{revtex4}
\usepackage{graphicx}
\usepackage{amsmath}
\begin{document}

\title{Wilson-Kadanoff Renormalization Group in Higher 
Orders: One-Dimensional g-ology Model as an Example }
\author{Gennady Y. Chitov}
\altaffiliation{Present address: Department 7.1-Theoretical Physics,
University of the Saarland, Saarbr\"ucken 66041, Germany}
\affiliation{Centre de recherche sur les propri\'et\'es \'electoniques de
mat\'eriaux avanc\'es, Universit\'e de Sherbrooke, Sherbrooke,
Qu\'ebec, Canada J1K 2R1}
\affiliation{Physics Department, McGill University,
Montr{\'e}al, Qu\'ebec,  Canada H3A 2T8}
\author{Claude Bourbonnais}
\affiliation{Centre de recherche sur les propri\'et\'es \'electoniques de
mat\'eriaux avanc\'es, Universit\'e de Sherbrooke, Sherbrooke,
Qu\'ebec, Canada J1K 2R1}
\date{\today}
\begin{abstract}
We apply the standard Wilson-Kadanoff (WK) momentum-space Renormalization
Group (RG) scheme for the g-ology model of one-dimensional fermions.
By explicitly carrying out calculations at the two-loop level,
we show how the RG flow equations can be derived from the summation
of the cascades of contractions generated by the effective action's
mode elimination at each infinitesimal step of the WK procedure.
The rules for selecting these series of cascades appear naturally
as a consequence of the WK scheme ``on-shell'' kinematic constraints and
conservation laws. The relation between the present RG approach
and the field-theoretic schemes used in earlier related studies
is analysed. Generalizations for other models and/or higher
dimensions are formulated.
\end{abstract}
\maketitle
%
%
\section{Introduction}\label{Intro}
%
%
The Renormalization Group (RG) approach, in its most
enlightening formulation due to Wilson \cite{WK74,Wilson},
is the theory designed to handle fields (quantum or classical)
fluctuating over range of momentum (energy) scales.\footnote{Since
the Wilson RG theory was strongly motivated
by Kadanoff's more intuitive original approach of successive averaging of
the spin Hamiltonian in real space \cite{Kadanoff},
it is often called the Wilson-Kadanoff (WK) RG, the term we apply in
this study.}
In the momentum-space formulation, this range of scales where
fluctuations matter is bounded by some initial ultraviolet (UV) cutoff
$\Lambda_0$ (not necessarily finite) provided by the problem under
consideration, and  $\Lambda_0$ is the scale where the RG procedure of
the successive mode integration starts. For the most problems where RG
was successfully applied (e.g., phase transition, field theories) the
low-energy sector of interest resides near one point of the momentum
space which can always be mapped to the origin of that space.

In practical loopwise calculations, the original WK infinitesimal scheme
seems however to appear too involved beyond one-loop level, and, e.g.,
for the Wilson's original problem of phase transition in higher orders
various field-theoretic RG approaches were applied \cite{BGZ76}. In
field theory the cutoff $\Lambda_0$ is essentially considered as a
``troublesome'' parameter to get rid of, either by eventually taking the
limit $\Lambda_0 \to \infty$, or, e.g, by using another regularization
scheme, like the dimensional one where the cutoff is set to infinity from
the start.

Contrary to the WK scheme, which specifically stipulates that at each
RG step the momenta to be integrated out lie within infinitesimally narrow
shell in the momentum space, in the field-theoretic RG only overall
momentum conservation is taken care of, while separate momenta may lie
outside of the (hyper)sphere of radius $\Lambda$ (centered at the origin).
This situation is well known, and contributions from the states lying
outside of the action's phase space at the distance of several radii
$\Lambda$-- or, in other terms, violating phase space constraints by
$\mathcal{O}(\Lambda)$-- do not affect the leading terms of the RG flow.

However, in condensed matter problems the Wilsonian \textit{effective
action's scale} $\Lambda_0$ must be taken seriously. \textit{A priori} it
cannot be set to infinity or even to the largest scale provided by the
underlying microscopics, i.e., inverse lattice, interatomic, etc,
spacing. Even for a problem with the low-energy sector localized near
the origin of the momentum space, aforementioned violations of the phase
space constraints result in double counting of the
degrees of freedom lying outside of the Wilsonian phase space (i.e.,
the degrees of freedom which are already ``integrated out'' in some way
before we reached the Wilsonian scale from the microscopic scale).
This can affect the next-to-leading terms, as well as finite
renormalizations of the action's parameters.

For the important problem of fermions having
a Fermi surface -- that is a low-energy sector residing near a manifold in
the momentum space -- accounting for the modes within the phase space
\textit {only} during the RG procedure becomes a rather non-trivial task,
as was explained by Shankar \cite{Shankar94}.
Moreover, the error of violating the phase space
constraints can be of order $\mathcal{O}(K_F)$, where $K_F$ is some
characteristic size of the Fermi surface, even if the low-energy modes
we are integrating out reside within a narrow shell $\Lambda \ll K_F$.

Note that the answers for many of the most interesting problems of the
condensed-matter fermions lie in calculating the abovementioned
``next-to-leading terms'', where the differences between various RG
approaches matter. For instance, the problem of the Fermi-Liquid vs
non-Fermi-Liquid regimes of the normal state of the cuprates, from an
RG point of view is the difference in the behavior of the irrelevant
(self-energy) terms in both regimes. So the choice of the optimal
RG scheme and the control over the subleading terms
are the issues far from being academical.

The main claim we want to make from the considerations of this paper
is that for a condensed-matter problem cast in terms of the Wilsonian
effective action with a physically meaningful UV cutoff $\Lambda_0$, the
WK RG provides the only scheme definitely reliable in all orders of
the loopwise expansions, where all modes from the phase space are
integrated once and only once. (Strictly speaking, there are no new
results in the RG equations we derive below, since for the $g$-ology
model they have been known for a long time, either from field-theoretical
RG scheme \cite{Kimura75,Solyom79} or more recent WK RG
formulations \cite{Bourbon91}). We address the issues of how to
practically employ the WK RG in higher orders, to which technically
simpler scheme the WK RG can be reduced and to what extend, by
studying the $g$-ology model of 1D fermions.

The model we consider is the simplest case of the condensed-matter
fermions where the Fermi surface manifold is just a set of two points.
By going deeper into a recent suggestion to this problem \cite{Bourbon02},
we apply a WK RG transformation assuring that at
each step of the procedure all momenta lie within the infinitesimal
shell of integration and satisfy the conservation laws, so there are no
contributions from the extra states nor double counting.
We show how the higher order renormalizations of the effective action's
parameters appear from the summation of the cascades of contractions
generated at each step of the WK procedure. We don't go beyond the
two-loop level, which is apparently the optimal one for feasible
calculations. At this level, we show that the summations over the
cascades in the WK scheme is equivalent (in the fixed-point limit of the
decreasing cutoff $\Lambda \to 0$) to applying the cutoff regularization
for the field-theoretic RG \textit{with the extra states excluded from
the phase space}.
This result, to the best of our knowledge, was not explicitly shown before.
Considerable simplifications for our analysis came from introducing the
chiral matrix formalism. This is a na\"ive (for the simple 1D case) first
attempt to build into the technique the disentanglement of geometry
(characterized by the ``large'' rigid scale $K_F$) and the low-energy
sector \textit{per se} (characterized by the decreasing scale $\Lambda$),
and to map the latter to the same origin. This mapping makes the exclusion
of the extra states particularly simple.

The rest of this paper is organized as follows. Sections
\ref{Model}-\ref{KWRG} are introductory. Section \ref{Model} defines the
effective action. In Section \ref{1DCEA} the chiral matrix formalism
is introduced. In Section \ref{KWRG} we explain the terms
and notations of the Wilson-Kadanoff RG as it is applied in this paper.
In Sections \ref{Cas}-\ref{2LC} we explain the procedure and give the
results for the two-loop self-energy calculations, and in
Sections \ref{WKFT} we establish the relationship between the
field-theoretic RG and the present
approach. Section \ref{Coupl} contains the two-loop results for the
couplings flow equations. The conclusions and generalizations for other
models and/or higher dimensions are presented in the final
Section \ref{Concl}.
%
%
\section{Model and Formalism}\label{MaF}
%
%
\subsection{$SU(N)$-invariant $d$-dimensional $\psi^4$-action}\label{Model}
We treat $d$-dimensional interacting fermions at finite temperature in the
standard path integral Grassmannian formalism \cite{Negele88}.
The $SU(N)$-invariant fermionic $\psi^4$-action along with the related
quantities and notations of this subsection
have been introduced earlier \cite{GD95}, and we recall these here in order
to make the paper self-contained.
The partition function is
\begin{equation}
\label{Z}
{\mathcal Z} =\int {\mathcal D}\bar\psi{\mathcal D}\psi~e^{S_0 + S_{\text{int}}}
\end{equation}
The free part of the action is
\begin{equation}
\label{S0}
S_0 = \int_{({\textbf 1})}\bar\psi_{\alpha}({\textbf 1})\left[
i\omega_1+\mu-\epsilon({\textbf K}_1)\right]\psi_{\alpha}({\textbf 1})
\end{equation}
and $(d+1)$-dimensional vectors and integrals mean
\begin{subequations}
\label{notation}
\begin{eqnarray}
\int_{({\textbf i})}~~&\equiv&~~\frac{1}{\beta}\int
\frac{d{\textbf K}_i}{(2\pi)^d}
\sum_{\omega_i}\\
({\textbf i}) ~~&\equiv&~~ ({\textbf K}_i,\omega_i)
\end{eqnarray}
\end{subequations}
where $\beta$ is the inverse temperature, $\mu$ the chemical potential,
$\omega_i$ the fermion Matsubara frequencies and $\psi_{\alpha}({\textbf i})$ a
$N$-component Grassmann field with a spin (flavor, if $N \neq 2$) index
$\alpha$. Summation over repeated indices is implicit throughout this
paper. We set $k_B=1$ and $\hbar =1$.
The $SU(N)$-invariant quartic interaction is
\begin{equation}
\label{Sint}
S_{\textrm{int}} = -\frac14\int_{({\textbf 1},{\textbf 2},{\textbf 3},{\textbf 4})}
\bar\psi_{\alpha}({\textbf 1})\bar\psi_{\beta}({\textbf 2})
\psi_{\gamma}({\textbf 3})\psi_{\varepsilon}({\textbf 4})
U^{\alpha\beta}_{\gamma\varepsilon}({\textbf 1},{\textbf 2};{\textbf 3},{\textbf 4})
\delta^{(d+1)}({\textbf 1}+{\textbf 2}-{\textbf 3}-{\textbf 4})
\end{equation}
Here the conservation of energy and momentum is enforced by the symbolic delta
function
\begin{equation}
\label{Delta}
\delta^{(d+1)}({\textbf 1}+{\textbf 2}-{\textbf 3}-{\textbf 4})\equiv\beta(2\pi)^d
\delta({\textbf K}_1+{\textbf K}_2-{\textbf K}_3-{\textbf K}_4)
\delta_{\omega_1+\omega_2-\omega_3-\omega_4,0}
\end{equation}
Note that for the lattice fermions the delta-function above conserves
momenta up to an inverse lattice vector.
The physical case of the spin-$\frac12$ electrons is recovered by setting
$N=2$. We decompose  the interaction by factorizing  its symmetric and
antisymmetric frequency-momentum- and spin- dependent parts as
\begin{equation}
\label{potS}
U^{\alpha\beta}_{\gamma\varepsilon} = U^A I^{\alpha\beta}_{\gamma\varepsilon}
+ U^S T^{\alpha\beta}_{\gamma\varepsilon}
\end{equation}
where the functions $U^{S/A}$ are symmetric/antisymmetric under exchange of the
left or right pairs of their variables. Two operators $\hat I$ and $\hat T$,
which are  respectively symmetric and
antisymmetric in the spin  space, are defined as follows:
\begin{subequations}
\label{base}
\begin{eqnarray}
I^{\alpha\beta}_{\gamma\varepsilon}~&\equiv&~
\delta_{\alpha\varepsilon}\delta_{\beta\gamma} +
\delta_{\alpha\gamma}\delta_{\beta\varepsilon}\\
T^{\alpha\beta}_{\gamma\varepsilon}~&\equiv&~
\delta_{\alpha\varepsilon}\delta_{\beta\gamma} -
\delta_{\alpha\gamma}\delta_{\beta\varepsilon}
\end{eqnarray}
\end{subequations}
The one-particle-irreducible (1PI) four-point vertex $\hat \Gamma$,
defined in the standard way, comprises two components
($\Gamma^A$, $\Gamma^S$) and can be written in the same manner  as
(\ref{potS}). Along with such representation showing
explicitly its antisymmetry, there is another one which
separates the vertex into charge and spin components via
\begin{equation}
\label{Vrep}
\hat \Gamma \mapsto
\Gamma^{\alpha \beta}_{\gamma \varepsilon}=
\Gamma^A I^{\alpha \beta}_{\gamma \varepsilon}+
\Gamma^S T^{\alpha \beta}_{\gamma \varepsilon}=
-\frac{1}{N}
\Gamma_{\text{c}}\delta_{\alpha \gamma}\delta_{\beta \varepsilon}
-\frac12 \Gamma_{\textrm{s}}
\lambda^a_{\alpha \gamma}\lambda^a_{\beta \varepsilon}
\end{equation}
The latter representation is often more convenient since charge and spin
components decouple in many practical calculations.
$\hat \lambda^a$ ($a=1,...,N^2-1$) in Eq.(\ref{Vrep})
are Hermitian traceless generators of the
$SU(N)$ group, coinciding with the Pauli matrices for $N=2$, and
normalized such that
\begin{equation}
\label{LamNor}
{\textrm{tr}}(\hat \lambda^a\hat \lambda^b)=2\delta^{ab}
\end{equation}
They also satisfy the following Fierz identity:
\begin{equation}
\label{Fierz}
\lambda^a_{\alpha \beta }\lambda^a_{\gamma \varepsilon}=
2 \Big( \delta_{\alpha \varepsilon}\delta_{\beta \gamma}
- \frac{1}{N} \delta_{\alpha \beta } \delta_{\gamma \varepsilon} \Big)
\end{equation}
The components of the vertex in different representations are related as
\begin{subequations}
\label{ABdef}
\begin{eqnarray}
\Gamma_{\textrm{c}} &=& (N-1) \Gamma^S-(N+1) \Gamma^A \\
\Gamma_{\textrm{s}}  &=& - \Gamma^S- \Gamma^A
\end{eqnarray}
\end{subequations}
%
%
%
\subsection{Chiral matrix formalism for the 1D effective action}\label{1DCEA}
The construction of the low-energy effective fermionic action as a simplified
form of the generic $\psi^4$-action (\ref{S0},\ref{Sint}) reduces in 1D
essentially to the three key steps: extraction of a finite set
of (marginal) couplings; decomposition of the physical electron $\psi$-field
into two chiral fields ``near'' left and right Fermi points; linearization
of the one-particle spectrum $\epsilon(K)$ around these points. There exists
an extensive literature on the derivation of the 1D effective action models
(Tomonaga-Luttinger, $g$-ology), and on many subtleties it involves (see,
e.g., Ref.[\onlinecite{Voit95}] and more references therein). So below we
simply present notations, approximations and highlight the main points.
The chiral matrix representation
we derive here is in fact only a way to re-write the well-known
$g$-ology model. (For details see the next subsection.) However this matrix
formalism and related to it diagrammatics help to facilitate bookkeeping of
diagrams and handling their contributions to different couplings, and,
eventually, to generalize the $g$-ology results for other cases.

We decompose 1D momenta as
\begin{equation}
\label{1DK}
K_i \equiv p_i K_F +k_i
\end{equation}
where $p_i= sign(K_i)=\pm 1$ corresponds to the right (R) /left (L) Fermi
points. In the quartic interaction function
$\hat U({\textbf 1},{\textbf 2};{\textbf 3},{\textbf 4})$ (\ref{potS})
we retain only its dependence on the
momenta at the Fermi points, neglecting that on $k_i$-s and
frequencies \cite{Dzyaloshinskii72}.
The $SU(N)$ and exchange symmetry constraints leave us with four independent
marginal couplings:
\begin{subequations}
\label{4coupl}
\begin{eqnarray}
g^A~&\equiv&~U^A(LR;LR)=-U^A(RL;LR)=-U^A(LR;RL) \\
g^S~&\equiv&~U^S(LR;LR)=\phantom{-} U^S(RL;LR)=\phantom{-} U^S(LR;RL) \\
g_4~&\equiv&~U^S(LL;LL)=U^S(RR;RR) \\
g_3~&\equiv&~U^S(LL;RR)=U^S(RR;LL)
\end{eqnarray}
\end{subequations}
Note that the Umklapp coupling $g_3$ is operative only in the half-filled
case when $4K_F=2 \pi$ (we set the lattice constant $a=1$). For the formalism
we develop below this case is
considered as the most general one, while the non-commensurate filling
can be recovered by setting $g_3=0$.

We introduce left (right) chiral Grassmannian fields as:
\begin{equation}
\label{ChirF}
\psi_{\alpha}(\mp K_F +k_i, \omega_i)=\psi_{\alpha}^{L/R}(k_i, \omega_i)
\end{equation}
Small vectors
\begin{equation}
\label{ksm}
k_i \in [-\Lambda_0, \Lambda_0]
\end{equation}
are restricted by the bare
momentum cutoff $\Lambda_0$ of the effective action.
Then the effective quartic interaction [cf. Eq.(\ref{Sint})] can be
represented as
\begin{equation}
\label{Seffint}
S^{(4)}_{\text{eff}} = -\frac14\int_{({ 1},{ 2},{ 3},{ 4})}
{\mathcal L}^{(4)}_{\text{eff}}~
\delta^{(1+1)}({ 1}+{ 2}-{ 3}-{ 4})
\end{equation}
where the transition from bold to thin variables corresponds to
$K_i \mapsto k_i$ [cf. notations (\ref{notation},\ref{Delta},\ref{1DK})],
and
\begin{eqnarray}
\label{Lint}
{\mathcal L}^{(4)}_{\text{eff}} &=&
\bar\psi_{\alpha}^L(1)\bar\psi_{\beta}^L(2)
\psi_{\gamma}^L(3)\psi_{\varepsilon}^L(4) [~~~0 \cdot
I^{\alpha\beta}_{\gamma\varepsilon}
+ g_4 \cdot  T^{\alpha\beta}_{\gamma\varepsilon}] + (L \leftrightarrow R)
\nonumber \\
&+&
\bar\psi_{\alpha}^L(1)\bar\psi_{\beta}^L(2)
\psi_{\gamma}^R(3)\psi_{\varepsilon}^R(4) [~~~0 \cdot
I^{\alpha\beta}_{\gamma\varepsilon}
+ g_3 \cdot  T^{\alpha\beta}_{\gamma\varepsilon}] + (L \leftrightarrow R)
\nonumber \\
&+&
\bar\psi_{\alpha}^L(1)\bar\psi_{\beta}^R(2)
\psi_{\gamma}^L(3)\psi_{\varepsilon}^R(4) [~~g^A \cdot
I^{\alpha\beta}_{\gamma\varepsilon}
+ g^S \cdot  T^{\alpha\beta}_{\gamma\varepsilon}] + (L \leftrightarrow R)
\nonumber \\
&+&
\bar\psi_{\alpha}^L(1)\bar\psi_{\beta}^R(2)
\psi_{\gamma}^R(3)\psi_{\varepsilon}^L(4) [-g^A \cdot
I^{\alpha\beta}_{\gamma\varepsilon}
+ g^S \cdot  T^{\alpha\beta}_{\gamma\varepsilon}] + (L \leftrightarrow R)
\end{eqnarray}
Note that the delta-function in (\ref{Seffint}), contrary to its ``bold''
counterpart (\ref{Delta}), conserves the small momenta $k_i$ exactly, since
the lattice effects at half-filling ($g_3$-terms) are explicitly taken into
account in (\ref{Lint}).

Let us now introduce four $2 \times 2$ matrices $\hat t^i$ such that $\hat t^0$
is the unit matrix, and $\hat t^i$ for $i=1,2,3$ correspond to the Pauli
matrices. Such distinct notation helps to avoid confusion, since $\hat t^i$
operates in the chiral, and not in the spin space. We attribute an
extra chiral index to the Grassmannian field as $L \mapsto 1$, $R \mapsto 2$.
Then four  terms in the last two lines of Eq.(\ref{Lint}) can be written as:
\begin{equation}
\label{Lfb}
{\mathcal L}_{\text{fb}}=
[-g^A \cdot I^{\alpha\beta}_{\gamma\varepsilon} \cdot
t^2_{\alpha' \beta'} t^2_{\gamma' \varepsilon'}
+ g^S \cdot  T^{\alpha\beta}_{\gamma\varepsilon} \cdot
t^1_{\alpha' \beta'} t^1_{\gamma' \varepsilon'} ]~
\bar\psi_{\alpha}^{\alpha'}(1)\bar\psi_{\beta}^{\beta'}(2)
\psi_{\gamma}^{\gamma'}(3)\psi_{\varepsilon}^{\varepsilon'}(4)
\end{equation}
while the first two lines in that equation give
\begin{equation}
\label{Luc}
{\mathcal L}_{\text{uc}}=
T^{\alpha\beta}_{\gamma\varepsilon} \cdot [g_p \cdot
t^0_{\alpha'\beta'} t^0_{\gamma' \varepsilon'} + g_m \cdot
t^3_{\alpha' \beta'} t^3_{\gamma' \varepsilon'} ]~
\bar\psi_{\alpha}^{\alpha'}(1)\bar\psi_{\beta}^{\beta'}(2)
\psi_{\gamma}^{\gamma'}(3)\psi_{\varepsilon}^{\varepsilon'}(4)
\end{equation}
where
\begin{equation}
\label{gpm}
g_{p/m} \equiv \frac12 (g_4 \pm g_3)
\end{equation}
Thus
\begin{equation}
\label{Lsum}
{\mathcal L}^{(4)}_{\text{eff}}={\mathcal L}_{\text{fb}} +{\mathcal L}_{\text{uc}}
\end{equation}
The one-particle Green's function is spin-independent in the model we
consider, but it depends on the chiral index:
\begin{equation}
\label{GFdef}
- \langle  \psi^{\alpha}_{\gamma}(1)
\bar \psi^{\beta}_{\varepsilon}(2) \rangle
=G_{\alpha}(1) \delta_{\alpha \beta} \delta_{\gamma \varepsilon}
\delta^{(1+1)}(1- 2)
\end{equation}
The linearized bare Green's function is given by
\begin{equation}
\label{GFbare}
G_{0,\alpha}^{-1}(k_n, \omega_n)= i \omega_n +\mu- \epsilon(p_{\alpha}K_F+k_n)
\approx  i \omega_n -p_{\alpha}v_F k_n
\end{equation}
where
\begin{equation}
\label{palf}
p_{\alpha}= 2(\alpha-1)-1
\end{equation}
Now we can define the 1PI four-point vertex $\hat \Gamma$ in this formalism.
We will be interested in a particular
limiting form of this vertex when its dependence on the external momenta lying
at the Fermi points (parametrized by the chiral indices) is
retained only, while ($k_n, \omega_n$) are discarded. In the lowest order
$\hat \Gamma^{(0)}$ can be read off from ${\mathcal L}^{(4)}_{\text{eff}}$
(\ref{Lfb},\ref{Luc},\ref{Lsum}).
As follows from the group-theory arguments,
this structure of the  vertex is preserved by interaction
${\mathcal L}^{(4)}_{\text{eff}}$: the latter results in only the renormalization of
couplings, but no new couplings are generated. Thus, the following operator
expansion with four couplings holds non-perturbatively:
\begin{equation}
\label{Gamg}
\hat \Gamma= -g^A~ \hat I \otimes \hat t^2 \otimes \hat t^2+
\hat T \otimes [g^S~ \hat t^1 \otimes \hat t^1 + g_p ~\hat t^0 \otimes \hat t^0
+g_m ~\hat t^3 \otimes \hat t^3]
\end{equation}
In order to keep the notations compact, we do not change them in (\ref{Gamg}),
however couplings are renormalized by interaction comparatively
to their bare values (\ref{4coupl}). To  the same end we will not indicate
explicitly the chiral and spin indices from now on (unless there is a possibility
of confusion), incorporating them into the ``thin variables''. So we imply
\begin{equation}
\label{compact}
{\mathcal L}_{\text{fb}} +{\mathcal L}_{\text{uc}} \mapsto ~
\text{[cf.~Eqs.(\ref{Lfb},\ref{Luc})]}~ \mapsto
\Gamma^{(0)}(12;34)\bar\psi(1)\bar\psi(2)\psi(3)\psi(4)
\end{equation}
Comparison of Eqs.(\ref{Lfb},\ref{Luc},\ref{compact}) with the operator  form
(\ref{Gamg}) clarifies how we attribute the chiral and spin indices for
the matrix elements of $\hat \Gamma$. Note the explicit antisymmetry of the
vertex $\Gamma(12;34)$ under exchange of the left or right pairs of variables
in the representation (\ref{Gamg}).

We can recover couplings from the vertex by appropriate convolutions, i.e.,
\begin{subequations}
\label{cps}
\begin{eqnarray}
g^A &=& -\frac{1}{48} \Big( t^2_{21}\cdot \Gamma(12;34) \cdot I^{43}_{21}
\cdot t^2_{43} \Big) \\
g_p &=& \phantom{-} \frac{1}{16} \Big( t^0_{21} \cdot \Gamma(12;34) \cdot
T^{43}_{21} \cdot t^0_{43} \Big)
\end{eqnarray}
\end{subequations}
and other two couplings are obtained from Eq.(\ref{cps}b) by obvious
substitutions, namely, $g^S:~\hat t^0 \mapsto \hat t^1$;
$g_m:~\hat t^0 \mapsto \hat t^3$.
%
%
%
%
\subsection{Contact with the $g$-ology model}\label{geol}
Here we want to make a contact between the chiral matrix formalism
and more commonly used $g$-ology language (see, e.g.,
Ref.[\onlinecite{Solyom79}]). No surprise that this subsection's formal
manipulations with the fully (anti)symmetrized ``chiral'' action
amount essentially to reducing it to the non-symmetrized form. The latter
is usually used as a departure point of various $g$-ology models.

Let us show first how the charge- and spin-operator
interaction  action, involving forward and backscattering couplings called
$g_2$, $g_1$ respectively in the $g$-ology nomenclature, can be recovered
from the part ${\mathcal L}_{\text{fb}}$ (\ref{Lfb}) of the effective quartic
interaction (\ref{Lsum}). ${\mathcal L}_{\text{fb}}$ can be split in two terms
by the following projection in the chiral space:
\begin{eqnarray}
\label{proj}
{\mathcal L}_{\text{fb}} &=& \Big( [1-t^0_{\alpha \gamma} t^0_{\beta \varepsilon}]
+t^0_{\alpha \gamma} t^0_{\beta \varepsilon} \Big)
(-g^A \hat I ~t^2_{\alpha \beta} t^2_{\gamma \varepsilon} +
  g^S \hat T ~t^1_{\alpha \beta} t^1_{\gamma \varepsilon})[\psi^4]_N
\nonumber \\
&\equiv& {\mathcal L}_{\text{cs1}}+{\mathcal L}_{\text{cs2}}
\end{eqnarray}
Above we indicate explicitly the chiral indices only, and $[\psi^4]_N$
stands for the product of four $\psi$-fields ``normally-ordered'' as on the
r.h.s. of Eq.(\ref{Lfb}).
Using the Fierz identity (\ref{Fierz}) for the $SU(2)$ chiral operators
$t^i$, one can prove that
\begin{eqnarray}
\label{Lcs1S}
{\mathcal L}_{\text{cs1}} &\equiv&
(1-t^0_{\alpha \gamma} t^0_{\beta \varepsilon})
(-g^A \hat I~ t^2_{\alpha \beta} t^2_{\gamma \varepsilon} +
  g^S \hat T~ t^1_{\alpha \beta} t^1_{\gamma \varepsilon})[\psi^4]_N
\nonumber \\
&=& \frac12 (t^1_{\alpha \beta} t^1_{\gamma \varepsilon}+
t^2_{\alpha \beta} t^2_{\gamma \varepsilon})
(-g^A \hat I+g^S \hat T)[\psi^4]_N
\end{eqnarray}
It can be also shown that the above expression reduces  to
\begin{equation}
\label{Lcs1SF}
{\mathcal L}_{\text{cs1}}=t^1_{\alpha \beta}(-g^A \hat I+g^S \hat T)
\bar\psi^{\alpha}(1)\bar\psi^{\beta}(2)
\psi^{\beta}(3)\psi^{\alpha}(4)
\end{equation}
In the same manner  we find
\begin{equation}
\label{Lcs2SF}
{\mathcal L}_{\text{cs2}}=t^1_{\alpha \beta}(g^A \hat I+g^S \hat T)
\bar\psi^{\alpha}(1)\bar\psi^{\beta}(2)
\psi^{\alpha}(3)\psi^{\beta}(4)
\end{equation}
Using the symmetry properties of the spin operators and changing
the dummy variables, it is easy to see that the term
${\mathcal L}_{\text{cs1}}$ results in the same expression for the action
[cf. Eq.(\ref{Seffint})] as ${\mathcal L}_{\text{cs2}}$, so
$S_{\text{fb}}=2S_{\text{cs2}}$. Applying then the Fierz identity (\ref{Fierz})
[cf. also definitions (\ref{base})] for the operators $\hat I, \hat T$
in case of $SU(2)$,
and using the conservation law for momentum-frequency such that
$3=1+Q$, $4=2-Q$, we get
\begin{equation}
\label{Sfb}
S_{\text{fb}}=
- \frac14 t^1_{\alpha \beta} \int_Q
\Big(g_c \rho^{\alpha}(Q)\rho^{\beta}(-Q)+
4 g_s S^{\alpha}_a(Q)S^{\beta}_a(-Q) \Big)
\end{equation}
where we introduced the operators of the chiral (i.e., left/right) charge
and of the chiral $a$-component of spin as
\begin{subequations}
\label{SCop}
\begin{eqnarray}
\rho^{\alpha}(Q) &=& \int_1
\bar\psi^{\alpha}_{\beta}(1)\psi^{\alpha}_{\beta}(1+Q) \\
S^{\alpha}_a(Q) &=& \frac12 \int_1
\bar\psi^{\alpha}_{\beta}(1) \lambda^a_{\beta \gamma}
\psi^{\alpha}_{\gamma}(1+Q)
\end{eqnarray}
\end{subequations}
Note that in the above formulas there are no sums over $\alpha$. We have
also defined the charge and spin couplings:
\begin{subequations}
\label{gSC}
\begin{eqnarray}
g_c &=& \phantom{-}g^S-3 g^A \\
g_s &=& -g^S- g^A
\end{eqnarray}
\end{subequations}
It is worth noting that although there is no one-to-one correspondence
between 1D couplings (\ref{4coupl}a,b) and the vertex components $\Gamma^A,
\Gamma^S$ [the latter in 1D are rather products of corresponding couplings
and operators in the chiral space, as can be seen from comparison of
Eqs.(\ref{Vrep},\ref{Gamg})], the definitions for the components
which couple  separated charge or spin modes (\ref{ABdef}) hold for the
chiral ones (\ref{gSC}). It is also useful to give the relationship between
our constants and the standard (dimensionful) $g$-ology couplings:
$g_c=2g_2-g_1$, $g_s=-g_1$.

The term $S_{\text{uc}}$ of the total quatric interaction generated
by the contact coupling at the same Fermi point ($g_4$) and the Umklapp
coupling ($g_3$) can be written as
\begin{eqnarray}
\label{Suc}
S_{\text{uc}} = &-& \frac{g_4}{4\cdot 2} \int_Q
\Big(\rho^{\alpha}(Q)\rho^{\alpha}(-Q)-
 4 S^{\alpha}_a(Q)S^{\alpha}_a(-Q) \Big) \nonumber \\
&-& \frac{g_3}{4\cdot 2} t^1_{\alpha \beta}  \int_{1,2,Q}
\Big( \delta_{\gamma \nu} \delta_{\sigma  \varepsilon}
- \lambda^a_{\gamma \nu} \lambda^a_{\sigma \varepsilon} \Big)
\bar\psi_{\gamma}^{\alpha}(1) \psi_{\nu}^{\beta}(1+Q) \cdot
\bar \psi_{\sigma}^{\alpha}(2)\psi_{\varepsilon}^{\beta}(2-Q)
\end{eqnarray}
The way we write the Umklapp contribution is to reinstate the
known fact that since $g_3$ couples fields with different
chiralities, it cannot be represented in terms of the local
chiral charge or spin operators.
%
%
%
%
\subsection{Wilson-Kadanoff renormalization group for the chiral effective
action}\label{KWRG}
In this study we apply the original Wilson-Kadanoff momentum-space
Renormalization Group scheme \cite{WK74,Wilson} to the effective action
with the quartic interaction containing four coupling constants.
This interaction in the chiral formalism is given by
Eqs.~(\ref{Gamg},\ref{compact}). In that formalism all the
constraints coming from the original two-point geometry of the
Fermi surface, and consequently, occurrence  of the two types
(left/right) of fermion fields and of Green's functions, are
taken care of by the appropriate summations over chiral indices.
In applying the RG procedure we successively integrate out the
Grassmannian fields in the phase space of the small momenta $k_i$
(\ref{1DK}). The latter, like in the quantum or classical
$\varphi^4$-theories, can be seen now as having common origin
towards which we eliminate modes in the $k$-space.

We start with the effective action (\ref{Gamg},\ref{compact}) having
momenta restricted by the initial (bare) ultra-violet (UV) cutoff
$\Lambda_0$ (\ref{ksm}). After $N$ steps of the mode elimination
($N \gg 1$), we end up with the cutoff $\Lambda$ chosen such that:
\begin{equation}
\label{Cutoffs}
\frac{\Lambda}{\Lambda_0} \ll 1, ~~\frac{v_F \Lambda}{T} \gg 1
\end{equation}
(We choose the RG scheme without rescaling of the cutoff.) Each
step consists of mode integration within the infinitesimal shell
$\Delta$ in the  $k$-space, where
\begin{equation}
\label{DeltaD}
\Delta \equiv \frac{\Lambda_0-\Lambda}{N}
\end{equation}
(the equidistant step is chosen for simplicity only) such that after n steps
the UV cutoff $\Lambda_n$ is
\begin{equation}
\label{Lambdan} \Lambda_n=\Lambda_0-n \Delta, ~~1 \leq n \leq N~~
(\Lambda_N \equiv \Lambda)
\end{equation}
We define the $n$-th shell in the momentum space ${\mathcal S}_n$ as
\begin{equation}
\label{Sn}
 {\mathcal S}_n \equiv [-\Lambda_{n-1}, -\Lambda_n] \cup [\Lambda_n,
\Lambda_{n-1}] \equiv {\mathcal S}_n^- \cup {\mathcal S}_n^+
\end{equation}
In the final results given in the following sections we will
smoothly take continuous limit $\Delta \to 0$ ($N \to \infty$)
such that relationship (\ref{DeltaD}) holds.
Then the lowering cutoff $\Lambda$ will be
determined by the continuous RG flow parameter $l$ defined as:
\begin{equation}
\label{lpar} \Lambda(l)=\Lambda_0 {\text{e}}^{-l}
\end{equation}
As it was discussed in the closely related earlier study of the
$g$-ology model \cite{Bourbon02}, the straightforward application of
the Wilson-Kadanoff RG beyond one-loop level encounters some
difficulties. They stem from the constraint on the momenta to
be integrated out at the $n$-th step to lie within the shell
${\mathcal S}_n$, and on the total momentum flowing through a given
graph to conserve. It was shown in Ref.~[\onlinecite{Bourbon02}] that
the solution of this problem consistent with the logic of the WK
RG can be found when multi-loop graphs (the standard ones for the
field-theoretical RG) renormalizing action's vertices are
obtained through cascades of contractions at different steps of
the RG procedure. In this study we refine the ``cascade scheme''
of the RG calculations put forward in Ref.~[\onlinecite{Bourbon02}]
for the $g$-ology model. We show that the aforementioned
constraints on the momenta can be cast in terms of the
``selection rules'' for the numbers of steps entering a given
cascade of contractions. Summations over all allowed types of
cascades provides the results particularly illuminating for
understanding the origin of the (logarithmic) equivalence of the
WK and field-theoretic RG versions.
%
%
%
\section{Self-Energy at the two-loop level}\label{Self}
%
%
%
\subsection{Cascades of contractions and kinematic constraints}\label{Cas}
We use the standard definition for the field renormalization constant
relating it to the self-energy (assuming that the analytical continuation from
Matsubara frequencies is done):
\begin{equation}
\label{Zf}
Z^{-1}=1- \frac{\partial \Sigma_{\alpha}(k, \omega)
 }{\partial \omega}\Big\vert^{\omega \to 0}_{k \to 0}
\end{equation}
It is easy to show that in our model $Z$ does not dependent on the
chirality index. The first correction to $Z$ comes from the
two-loop ``sunrise'' self-energy diagram. However, in the WK RG
scheme such a diagram if obtained in the second order by a
contraction of three legs of the four-leg interaction
(\ref{Seffint}) with itself, gives a vanishingly small
contribution. This is due to the momentum constraints discussed
above. (See also Ref.~[\onlinecite{Bourbon02}] for more details.) The
graphs we are looking for can come only from two types of
cascades of contractions generated by the mode elimination of the
original action (\ref{Seffint}): 3-step and 2-step cascades
depicted in Fig.~1. \textit{3-step cascade} (Fig.~1a) is generated
(in the second order over interaction) as follows: (i) at the
$n$-th step of the mode elimination the four-point vertex is
contracted with itself to generate the six-leg vertex; (ii) $m$
steps after [i.e. at the ($n+m$)-th step] two legs of this vertex
are contracted; (iii) at the ($n+m+k$)-th step two other legs are
contracted resulting in the final correction to the two-leg
vertex (i.e., one-particle Green's function). \textit{2-step
cascade} (Fig.~1b) is generated as follows: (i) at the $n$-th
step two pairs of legs of the four-point vertices are contracted;
(ii) at the ($n+m$)-th step two more legs are contracted. The
above description of the cascades was very schematic and
illustrative. We did not take care of the kinematics of diagrams
(that is why we did not indicate the momentum directions in the
graphs of Fig.~1), and the integers numerating the steps are
unrelated and free to take any values restricted only by the
total number of steps of the RG procedure, i.e., $1 \leq n \leq
n+m \leq n+m+k \leq N$.\footnote{Note that 2-step cascade or
``1-step cascade'' (i.e., simple contraction) can be considered
as special cases of a 3-step cascade if we allow the integers $m,k$
to take zero values.}
\label{nm}

To proceed with the further analysis of the pertinent cascades let
us consider the renormalization of the one-particle Green's
function at $k=\Lambda$. As the RG flows towards the fixed point
($\Lambda \to 0$) this quantity will provide us with the field
renormalization constant $Z$ (\ref{Zf}). For this purpose we
consider the terms generated by the two types of cascades
described above renormalizing the coefficient of the action's
Grassmannian binomial $\bar \psi(1) \psi(1)$.
Diagrammatically  they are shown in Fig.~2.  Analysis reveals that
there are three kinematically non-equivalent graphs ($a, b, c$ in
that figure) generated by the 3-step cascades and two graphs ($d,
e$) from 2-step cascades. We choose the external momentum
\begin{equation}
\label{k1} k_1=\Lambda_N \equiv \Lambda
\end{equation}
and the other external variables (chirality, spin, frequency) are
not specified at this point. For the further applications we
approximate
\begin{equation}
\label{LamL} \Lambda = L \Delta
\end{equation}
where L is an integer ($1 \ll L \ll N$). The error ${\mathcal
O}(\Delta)$ of the representation (\ref{LamL}) disappears when we
eventually take the limit $\Delta \to 0$.

Let us start with the kinematics of the first three graphs. We
find that for each of these graphs there are two solutions
compatible with the constraints imposed by the WK procedure and
the momentum conservation law. They are given in Table I.
\\[0.5cm]
Table I: \textit{Solutions for the momenta of the three
graphs in Fig.~2 compatible with the kinematic constraints.}
\\[0.3cm]
\begin{tabular}{|c|c|c|c|c|c|c|} \hline
Graph & a & a & b & b & c & c \\[0.8ex]
\hline
~~Solution~~ & I & II & I & II & I & II \\[0.8ex]
\hline
m'= & m & m-2L & m & m-2L & m & m-2L \\[0.8ex]
\hline
$k_2 \in$ & $~~{\mathcal S}_{N-m}^-~~$   & $~~{\mathcal S}_{N-m+2L}^+~~$
  & ${\mathcal S}_{n}^+$  & ${\mathcal S}_{n}^-$
  & ${\mathcal S}_{n+m}^-$ & ${\mathcal S}_{n+m}^+$ \\[0.8ex]
\hline
$k_3 \in$ & ${\mathcal S}_{n}^-$   & ${\mathcal S}_{n}^+$
  & $~~{\mathcal S}_{N-m}^+~~$  & $~~{\mathcal S}_{N-m+2L}^-~~$
  & $~~{\mathcal S}_{N-m}^+~~$ & $~~{\mathcal S}_{N-m+2L}^-~~$ \\[0.8ex]
\hline
$k_4 \in$ & ${\mathcal S}_{n+m}^+$   & ${\mathcal S}_{n+m}^-$
  & ${\mathcal S}_{n+m}^+$  & ${\mathcal S}_{n+m}^-$
  & ${\mathcal S}_{n}^-$ & ${\mathcal S}_{n}^+$ \\[0.8ex]
\hline
\end{tabular}
\\[0.5cm]
The ranges for the integers $n, m$ satisfying the solutions
I, II are:
\begin{subequations}
\label{nmranges}
\begin{eqnarray}
I~(a,b,c): ~&1& \leq n \leq N-2m;~~~~~ 1 \leq m \leq \frac{N}{2}  \\
II~(a,b,c):~&1& \leq n \leq N+2(L-m);~ 2L < m \leq \frac{N}{2}+L
\end{eqnarray}
\end{subequations}

Thus in order to obtain the total contribution of these
cascade-generated graphs to the self-energy we have to integrate
two independent momenta in each graph over available range (i.e.
two possible configurations I, II for each graph) and to sum over
all allowed types of cascades parameterized by ($n, m$) within the
ranges (\ref{nmranges}). (And of course, there are always
straightforward summations over other variables, i.e., chirality,
spin, frequency which take all available values.)

We have not shown in Table 1 available kinematic solutions
for the 2-step cascade graphs d, e, since their contribution is
negligible in comparison to that of a-c. Indeed, as can be shown
from the results below, the graphs d, e as well as a-c \textit{at a
fixed m} give the contributions ${\mathcal O}(\Delta / \Lambda)$ to
the constant $Z$. However the extra sum over $m$ renders the
latter logarithmic ${\mathcal O}(\ln \frac{\Lambda_0}{\Lambda})$,
while the former do not have this extra sum and thus strictly
disappear in the limit $\Delta \to 0$. For the same reason we
drop some extra solutions for the graphs a-c which occur only
under special relationships between $n$ and $m$ (i.e., there is
no double sum over $n, m$): they give vanishing
contributions.\footnote{In the light of the previous comment
(see the footnote on p.~\pageref{nm})
it is clear that special configurations (or ``lower-order'' cascades)
are in fact recovered in the double integrals appearing in the limit
$\Delta \to 0$ as some points or lines, i.e., manifolds of inferior
measure.}
%
%
%
%
\subsection{Two-loop calculations}\label{2LC}
Now we can proceed with the RG calculations via a straightforward
evaluation of the graphs from Fig.~2.  For certainty we fix the
external chiral index of those graphs to be 2, i.e. corresponding
to the right component of the Grassmannian field. Since graph's
contributions in the spin space are diagonal and spin-independent,
the external spin can be ``up'' or ``down''. After summation over
chiral and spin indices we obtain for the sunrise graph
contribution ${\text{F}}_+(1)$ (here plus stands to indicate our
choice of the external chiral index):
\begin{equation}
\label{Fplus}  {\text{F}}_+(1)= \int_{({ 2},{ 3},{ 4})} \Big[\frac12
\big( g_c^2+3g_s^2 \big) \cdot {\mathcal F}_+^{cs} + 2g_3^2 \cdot {\mathcal
F}_+^u + 2g_4^2 \cdot {\mathcal F}_+^4 \Big]
 ~\delta^{(1+1)}({ 1}+{ 2}-{ 3}-{ 4})
\end{equation}
where thin variables signify momenta and frequencies, i.e., the
only ones left, and
\begin{subequations}
\label{Fs}
\begin{eqnarray}
{\mathcal F}_+^{cs} ~&\equiv&~ G_-(2) \big[ G_-(3)G_+(4)+G_+(3)G_-(4)\big] \\
{\mathcal F}_+^u~&\equiv&~ G_+(2) G_-(3) G_-(4) \\
{\mathcal F}_+^4~&\equiv&~ G_+(2) G_+(3) G_+(4)
\end{eqnarray}
\end{subequations}
Note that until the integration ranges for the momenta $k_2, k_3,
k_4$ are specified, the same analytical expression ${\text{F}}_+(1)$
(\ref{Fplus}) corresponds to the \textit{each of graphs} depicted in
Fig.~2.

The following summation of Matsubara frequencies in (\ref{Fplus})
is straightforward. Due to conditions (\ref{Cutoffs}) we
approximate the occurring hyperbolic functions up to
exponentially small terms ${\mathcal O}(\exp(- \Lambda/T))$. With
this accuracy we drop all contributions from the term ${\mathcal
F}_+^4$. Analysis of the available kinematic configurations from
Table 1 for other terms shows that some of them are also
exponentially small, so we put
\begin{subequations}
\label{Fszero}
\begin{eqnarray}
{\mathcal F}_+^{cs} \big\vert_{bI,II}=0 \\
{\mathcal F}_+^u \big\vert_{aI,II}=0 \\
{\mathcal F}_+^u \big\vert_{cI,II}=0
\end{eqnarray}
\end{subequations}
Our calculations show that the term ${\mathcal F}_+^{cs}$ evaluated for
the configurations of the graph a (solutions I, II) and c (I, II)
gives four equal contributions to the self-energy. Analogously,
two contributions of ${\mathcal F}_+^u$ from b (I, II) are equal to
each other. In order to demonstrate how these calculations are
done, let us consider as an example ${\mathcal F}_+^{cs}$ for the
configuration (a, I). To obtain its total contribution to the
self-energy, we have to to integrate two independent momenta
within the ranges indicated in Table I for fixed $n, m$ and then
to sum over all available cascades (a, I) according to
(\ref{nmranges}a). This gives:
\footnote{Strictly speaking, couplings in the sunrise graphs depend
on the smallest shell number, i.e., n. By pulling them out of sums
and integrals, as should be clear from the content of this
section, we actually approximate them by their values at the
largest shell number (i.e., at $\Lambda$). This approximation
(called ``local'' in Ref.[\onlinecite{Bourbon02}]) is common for all RG
schemes, and in any case possible corrections to it lie beyond
the two-loop level.} \label{notegn}
\begin{equation}
\label{Siga1dis} \Sigma_+^{cs}(1)\Big\vert_{a1}= \frac{v_F}{16} \big(
g_c^2+3g_s^2 \big) \sum_{m=1}^{N/2} \sum_{n=1}^{N-2m}
\int_{-\Lambda-m \Delta}^{-\Lambda-m \Delta+\Delta} dk_2
\int_{\Lambda_n-m \Delta}^{\Lambda_n-m \Delta+\Delta}
\frac{dk_4}{a_+-k_2+k_4}
\end{equation}
where
\begin{equation}
\label{apl}
  \frac{1}{a_+ } \equiv 2 v_F G_{0,+}(1)
\end{equation}
In Eq.~(\ref{Siga1dis}) and in the following we use dimensionless
couplings absorbing the density of states factor in their
definition, i.e.,
\begin{equation}
\label{dlsc}
  \frac{g_{\sharp}}{\pi v_F} \mapsto g_{\sharp}
\end{equation}
In the limit $\Delta \to 0$ the sums above neatly combine into a
simple double integral
\begin{equation}
\label{Siga1int} \Sigma_+^{cs}(1)\Big\vert_{a1}= \frac{v_F}{16} \big(
g_c^2+3g_s^2 \big) \int_{-\frac12(\Lambda_0+\Lambda)}^{-\Lambda}
dk_2 \int_{-k_2}^{\Lambda_0+\Lambda+k_2} \frac{dk_4}{a_+-k_2+k_4}
\end{equation}
which gives
\begin{equation}
\label{dSiga1}
  \frac{\partial \Sigma_+^{cs}(0)}{\partial \omega} \Bigg\vert_{a1}
  =-\frac{g_c^2+3g_s^2}{4 \cdot 16} \big( \ln \frac{\Lambda_0}{
  \Lambda}+... \big)
\end{equation}
where the omitted terms $const +{\mathcal O}(\Lambda / \Lambda_0)$ are
irrelevant for the RG flow equations. Combining evaluated in the
same way other contributions and using the definitions
(\ref{lpar},\ref{Zf}), we obtain the two-loop RG equation
\begin{equation}
\label{ZRG}
  \frac{\partial \ln Z}{\partial l}= -\frac{1}{16}
  (g_c^2+3g_s^2+2g_3^2)
\end{equation}
This equation was first obtained in Ref.[\onlinecite{Kimura75}].
%
%
%
%
\subsection{Wilson-Kadanoff vs field-theoretic schemes}\label{WKFT}
In order to get more insights from the obtained results, let us
first rewrite some of the previous formulas. Using the transfer
momentum vector $q=k_2-k_4$ in Eq.(\ref{Siga1int}), we get
\begin{equation}
\label{Sigtr} \Sigma_+^{cs}(1)\Big\vert_{a1}= \frac{v_F}{16} \big(
g_c^2+3g_s^2 \big) \int_{-\frac12\Lambda_0}^{-\Lambda} dk
\int^{2k}_{-\Lambda_0} \frac{dq}{a_+-q}
\end{equation}
In the above equation we dropped the index of the vector $k_2$
and neglected terms $\Lambda \ll \Lambda_0$ in the limits of
integrals, since we are eventually interested only in the
$\Lambda \to 0$ limit. Other three non-zero
contributions [cf. Eq.(\ref{Fszero}a)] to the $cs$-term of the
self-energy after analogous variable changes can be reduced to
the form (\ref{Sigtr}) but with different integration ranges.
Combined together, these contributions can be written as:
\begin{equation}
\label{Sigs} \Sigma_+^{cs}(1)= \frac{v_F}{16} \big( g_c^2+3g_s^2
\big) \int_{-\frac12\Lambda_0}^{\frac12\Lambda_0}  dk \Theta(k^2
\geq \Lambda^2) \int^{\Lambda_0}_{-\Lambda_0}
\frac{dq \Theta(q^2 \geq 4k^2)}{a_+-q}
\end{equation}
where $\Theta$ is the event (Heaviside) function.

Thus, by applying the canonical Wilson-Kadanoff scheme where all
momenta in the diagrams lie strictly within an infinitesimal shell
at each step of the RG procedure, and by summing all pertinent
infinitesimal contributions through  singling out
kinematically-allowed cascades, we end up with results which could
have been obtained in the field-theoretic scheme. As one can see,
formula (\ref{Sigs}) for the self-energy resembles
the one derived in the RG scheme where momenta are let to run
freely within a whole band restricted only by the IR and UV
cutoffs.

To sharpen the last statement, let us derive now the self-energy
in the field-theoretic RG scheme with the cutoff regularization.
(Thorough description of various field-theoretic RG approaches can be
found, e.g., in Ref.[\onlinecite{Zinn99}].) Let us assume that we have
the same bare effective action as we used before. Since the theory is
logarithmical, we regularize divergent integrals by introducing
IR and UV  cutoffs. We assume that they satisfy the same
conditions as those for the WK effective action (\ref{Cutoffs}).
Since \textit{a priori} the cutoffs in the field-theoretic and WK
approaches are not the same, we will mark the former with
tildes.\footnote{For the IR cutoff $\tilde \Lambda$ to be the
smallest scale of the problem we will work with the theory in the
zero-temperature limit. For the following comparisons note that
the second condition (\ref{Cutoffs}) effectively brings the WK
RG results to the zero-temperature limit as well.}
Contrary to the WK scheme, in the
field-theoretic approach the self-energy contribution we are
looking for: \textit{(i)} comes from the \textit{only one} sunrise
graph as those depicted in Fig.~2; \textit{(ii)} independent momenta
in the graph are restricted by the model cutoffs only ;
\textit{(iii)} the graph has a combinatorial factor $1/2$. All the
summations before the final momentum integrations in this graph
lead to the same formulas as in the previous subsection (except
the overall factor $1/2$). We choose a couple of independent momenta
$k_i$ restricted by the cutoffs $\tilde \Lambda, \tilde \Lambda_0$.
It can be shown that the term of the self-energy involving charge and
spin couplings can be brought to the form (\ref{Sigs}) with
$\tilde \Lambda \mapsto \Lambda$, $\tilde \Lambda_0 \mapsto \Lambda_0$
up to ${\mathcal O} ( \tilde \Lambda / \tilde \Lambda_0)$.
With the same accuracy the self-energy Umklapp terms coincide in
the WK and field-theoretic approaches.
Thus both schemes result in the same leading $\ln (\Lambda_0
/\Lambda)$ term in the derivative of the self-energy with respect
to the frequency, and therefore to the same RG equation (\ref{ZRG}).

The relationship between the two RG schemes can be seen even in a simpler
way in the limit $\Lambda \to 0$, i.e., the fixed point limit we are always
interested in. Let us draw the regions of momenta integrations corresponding
to the solutions from Table I summed over available $n$ and $m$.
As a couple of independent momenta we choose
$(k_3,k_4)$. These regions are shown in Fig.~3. Let us remind also that
while showing kinematically non-equivalent cascades (a-c) in Fig.~2 we
have used the (anti)symmetry of the vertex, and, correspondingly, that of
graph's contributions under the exchange ($ 3 \leftrightarrow 4$).
So the sum of these three graphs can also be written as
$(a+b+c)= \frac12 (a+b+c+a'+b'+c')$, where the primed
graphs correspond to the above exchange. As one can see from Fig.~3
the regions for unprimed and primed solutions (the primed solutions are
obtained from their unprimed counterparts by the reflection over the main
diagonal), fill completely the hexagon shown in bold. The latter is nothing
but the whole phase space of the sunrise graph
\begin{equation}
\label{Srestr}
{\mathcal S}_{\text{restr}}:~|k_{i}| \leq \Lambda_0,~i=2,3,4;~k_1=0.
\end{equation}
Now assuming that the summations and integrations commute we can write
the total sunrise self-energy correction $\Sigma_{\text{sr}}^{\text{casc}} $
resulting from the sum of three kinematically distinct graphs 2a-c
over two solutions for each graph and over all available cascades
as a single term
\begin{equation}
\label{SigSR}
 \Sigma_{\text{sr}}^{\text{casc}} (1)=
 \frac12 \int_{({ 2},{ 3},{ 4});k_i \in {\mathcal S}_{\text{restr}}}
 \Gamma(12,34)G(2)G(3)G(4)\Gamma(34,12)
 ~\delta^{(1+1)}({ 1}+{ 2}-{ 3}-{ 4})
\end{equation}
Here we came back to the original thin variables comprising chiral and spin
indices, Matsubara frequencies, and momenta ($\omega_1,k_1 \to 0$).
Thus the self-energy correction has the same analytical form
as the one written in the cutoff field-theoretic approach. In the latter
however, after removing one momentum by the delta-function (e.g., $k_2$)
only two independent momenta are explicitly restricted by
$\Lambda_0$. This results in the contributions from the extra states
forbidden by the whole phase space constraints. These extra states lie in
the triangles in the first and third quadrants of the $(k_3,k_4)$-plane
(cf. Fig.~3). In the logarithmic theory as the one we are dealing with,
integration over these forbidden states may result in only the finite
terms to the self-energy correction, thus leaving the RG equation
(\ref{ZRG}) unaffected.\footnote{It is possible, (e.g., it happens for the
component $\Sigma^{\text{cs}}$) that summation over frequencies
bringing the Fermi functions into play, suppresses completely
the differences between two schemes coming from the extra states. For
$\Sigma^{\text{cs}}$ the contributions from the first and third quadrants
are completely eliminated by the Fermi functions [cf.
Eq.(\ref{Fszero}a)]. But generically there are no reasons for
that, and the Umklapp contribution provides a counterexample
for $\Sigma^{\text{cs}}$. Cf. Eqs.(\ref{Fszero}b,c).}

The phase space mapping of Fig.~3 holds in the fixed-point limit
$\Lambda \to 0$ at non-zero temperature as well. Thus the temperature
effects can be calculated from the UV cutoff RG scheme upon integration
of the all phase space restricted by the hexagon in
Fig.~3.\footnote{Note, however that the explicit form of the flow
equations, i.e., the Gell-Mann--Low functions, derived here in the limit
$T \to 0$, are modified at $T \neq 0$ when decreasing cutoff $\Lambda(l)$
reaches the scale provided by the temperature.}
As it should be clear, this is not equivalent either to calculation
within WK RG stopped at some $l^*$ when $v_F \Lambda(l^*) \sim T$,
or within the two-cutoffs RG with the IR cutoff provided by $T$:
such two ways of calculations give, strictly speaking, approximations
only.
%
%
\section{Couplings at the two-loop level }\label{Coupl}
%
%
After explaining most of technicalities in the previous section, we will be
brief in presenting the results for the RG coupling equations. Let us first
write the two-loop perturbative expansion for the four-point 1PI vertex.
It is depicted graphically in Fig.~4. (The tadpole self-energy corrections
are absorbed in the one-particle Green's functions renormalizations.)
We calculate the couplings from $\Gamma(12,34)$ at
\begin{equation}
\label{GamLim}
k_i=0, ~\omega_i=
\omega_{\circ}~(i=1,..4)
\end{equation}
where $\omega_{\circ}$ is the minimal Matsubara
fermionic frequency. The latter is set to zero after appropriate analytical
continuation. The couplings are extracted from $\Gamma$ by convolutions
[cf. Eqs.(\ref{cps}) and below].
The limits (\ref{GamLim}) considerably reduce the number of cascades to
take into account while calculating the renormalized couplings.

One-loop corrections are generated by a simple contraction shown in the
first line of Fig.~1b. After summations over the shell number these
contributions reduce to three one-loop graphs with the
signs and factors as in the pertubrative expansion in Fig.~4 and the loop
momentum $\Lambda \leq |k| \leq \Lambda_0$.

Pertinent double-loop graphs are generated by the 2-step cascade as
follows: at the $n$-th step a simple one loop is generated (cf. first line
of Fig.~1b) and at the $m$-th step the left (or right) pair of legs of this
graph is contracted with the four-leg vertex. The shell numbers $n$ and $m$
are independent and can take any value from 1 to $N$. The sums of these
cascades result in the three double-loop graphs (DL) as they stand in the
second line in Fig.~4 with the loop momenta restricted by $\Lambda \leq |k|
\leq \Lambda_0$.

The two-loop graphs combined in the groups (ZZ, ZC, CZ) in Fig.~4 are
generated by the 3-step cascades. For concreteness let us take the
first graph from the group ZZ. For each type of this loop configuration
with a fixed position of external variables (top left in Fig.~5) there are
six graphs corresponding to all possible types of the 3-step contractions
shown in this figure. Direct inspection shows that the sign and
combinatorial factor of all these six graphs are the same and coincide
with those of their counterpart in the perturbative expansion in Fig.~4.
Due to conditions (\ref{GamLim}) we retain only such cascades that the legs
of the vertex $\alpha$ are contracted at the same step. From the
conservation law the momenta $k_5,k_6$ are equal (opposite for the CZ
graphs), so the phase space constraints near the vertex $\alpha$ are
satisfied trivially. To find the allowed kinematic solutions from the
second independent momentum conservation (e.g., at the vertex $\beta$ of
this graph) we can use directly the results of the previous section.
Indeed, if we identify $k_2,k_8,k_5,k_7$ of the graphs in Fig.~5 with
$k_1,k_2,k_3,k_4$ of Fig.~2, respectively, we recover the situation
analysed previously for the self-energy graphs.
The kinematic solutions for the graphs of Fig.~5 can be taken from Table I
(in the limit $L=0$) as follows: $i \mapsto a$, $ii \mapsto c'$,
$iii \mapsto a'$, $iv \mapsto b'$, $v \mapsto c$, $vi \mapsto b$. Then
contributions $i-vi$ (cf. footnote on p.~\pageref{notegn}) can be combined
into a single graph (top left in Fig.~5) with two independent momenta lying
in the phase space shown in Fig.~3. Analogous conclusions can be reached
for other graphs of the groups ZZ, ZC, CZ. Thus, provided the stipulated
above phase space of each graph, the renormalization of the vertex
after mode elimination between $\Lambda$ and $\Lambda_0$ ($\Lambda \to 0$)
is given by Fig.~4.

The rest of the calculations is straightforward.
We introduce the renormalized (dimensionless) couplings as follows:
\begin{equation}
\label{gr}
g_{\sharp}=Z^2Z_{\sharp}^{-1} g_{\sharp}^{\text{(bare)}}
\end{equation}
The vertex renormalization constants $Z_{\sharp}^{-1}$ in terms of
$g_{\sharp}$ are found to be:
\begin{subequations}
\label{Zsharp}
\begin{eqnarray}
Z_c^{-1} &=& 1+\frac{g_3^2}{g_c}l +\frac18(g_c^2+3g_s^2-2g_3^2)l
 -g_3^2 \big(1-\frac{g_3^2}{g_c^2}\big)l^2 \\
Z_s^{-1} &=& 1+g_s l+\frac18(g_c^2-g_s^2+2g_3^2)l \\
Z_3^{-1} &=& 1+g_c l -\frac18(g_c^2-3 g_s^2)l + \frac12 (g_c^2-g_3^2)l^2
\end{eqnarray}
\end{subequations}
Combining the equations
\begin{equation}
\label{gMast}
\frac{d \ln g_{\sharp}}{dl}=2 \frac{\partial \ln Z }{ \partial l}
+\frac {\partial \ln Z_{\sharp}^{-1} }{ \partial l}+
\sum_{i=c,s,3}
g_i \frac{\partial \ln Z_{\sharp}^{-1} }{ \partial g_i}
 \frac{d \ln g_i }{ dl}
\end{equation}
with Eqs.(\ref{ZRG},\ref{Zsharp}) we obtain the RG flow
equations for couplings:
\begin{subequations}
\label{gRG}
\begin{eqnarray}
\frac{d g_c }{ dl} &=& g_3^2-\frac12 g_cg_3^2 \\
\frac{d g_s }{ dl} &=& g_s^2-\frac12 g_s^3 \\
\frac{d g_3 }{ dl} &=& g_3g_c -\frac14 g_3(g_3^2+g_c^2)
\end{eqnarray}
\end{subequations}
These equations were obtained earlier within the field-theoretic
RG analysis \cite{Kimura75}. Note that in the above equations we dropped
all the contributions from $g_4$ assuming that they can be absorbed
in the renormalization of Fermi velocities. (See, e.g.,
Ref.~[\onlinecite{Solyom79}] and more references therein.)

%
%
\section{Conclusions }\label{Concl}
%
The equivalence of the RG flow equations in the WK and field-theoretic
schemes at the two-loop level is proven explicitly for the 1D $g$-ology
model. To obtain the higher-order RG equations within the canonical
WK approach,  it is crucial to sum over all available cascades of
contaractions occurring at the infinitesimal steps of the WK procedure.
Moreover, in the fixed-point limit $\Lambda \to 0$ the diagrammatics
of the WK RG summed over all cascades is shown to be the same as that
of the field-theoretic RG with the cutoff regularization and restricted
phase space.

Generalizations for the cases of classical or quantum bosonic and
fermionic fields and higher dimensions are straightforward within the
developed technique, resulting in the same conclusion about the
diagrammatics, provided that the relevant phase space of those fields
(i.e., the low-energy sector) can be brought to the same origin.

If the latter condition can not be satisfied easily, like, e.g, for the
low-energy sector of higher dimensional condensed-matter
fermions possessing a Fermi surface, the proof of equivalence
is not a straightforward extension within the proposed approach,
and more explicit work is needed to demonstrate it.
However we conjecture that even in that case the summation over cascades
within the WK RG tantamount to applying the field-theoretic RG with
a (bare) UV cutoff for the low-energy sector and locally-imposed
phase-space constraints of the integrated momenta. (For the examples
of how these constraints may be realized, see the review by Shankar
[\onlinecite{Shankar94}].)

At the level of three loops and higher, different RG schemes
may result in different flow equations. Within the present approach,
it follows from the necessity to take into account the dependence of
running couplings on the shell number in higher orders (cf.
footnote on p.~\pageref{notegn}). So various cascade contributions
could not be ``compactified'' in a single ``field-theoretic'' graph as
it occurs at the two-loop level. The latter conclusion is hardly
surprising, since the differences in the Gell-Mann--Low functions
in higher orders are known to occur even within different versions
of the field-theoretic RG. (See, e.g, Ref.[\onlinecite{Zinn99}].
In the context of the 1D $g$-ology, see also Ref.[\onlinecite{Rezayi79}]
for the discussion of the bandwidth- and transfer-cutoff regularizations
in higher orders of RG.)

In condensed-matter problems, where the bare UV cutoff $\Lambda_0$
might be a meaningful scale and other scales (e.g., $T$, $K_F$)
may be present as well, the WK RG scheme should be
applied, as the one which is just an exact way to to integrate
the partition function \cite{WK74,Wilson}. The WK RG does not
rely on any assumptions of the scales involved, scaling/logarithmicity
of theory, etc, and includes contributions only from the states
lying within the phase space of the problem. This allows to control
the results (at least in principle) in all orders.
In practical terms, after summations over all cascades the WK scheme
asymptotically maps to the much easier to handle field-theoretic
RG (up to three loops) with forbidden states excluded.
We did not verify whether possible simplifications of the WK
scheme can be done in even higher orders,
but going beyond two-loop level is a formidable problem in any case.

%
\begin{acknowledgments}
G.Y.C. is grateful to M.E. Fisher and A.M. Tsvelik for helpful discussions,
and to McGill University for hospitality. C.B. thanks R. Wortis for
important remarks  at the early stage of this work.
This work is supported by NSERC (Canada), FCAR (Q\'ebec), and German
Science Foundation.
\end{acknowledgments}
%

\begin{figure*}
\includegraphics[width=09cm]{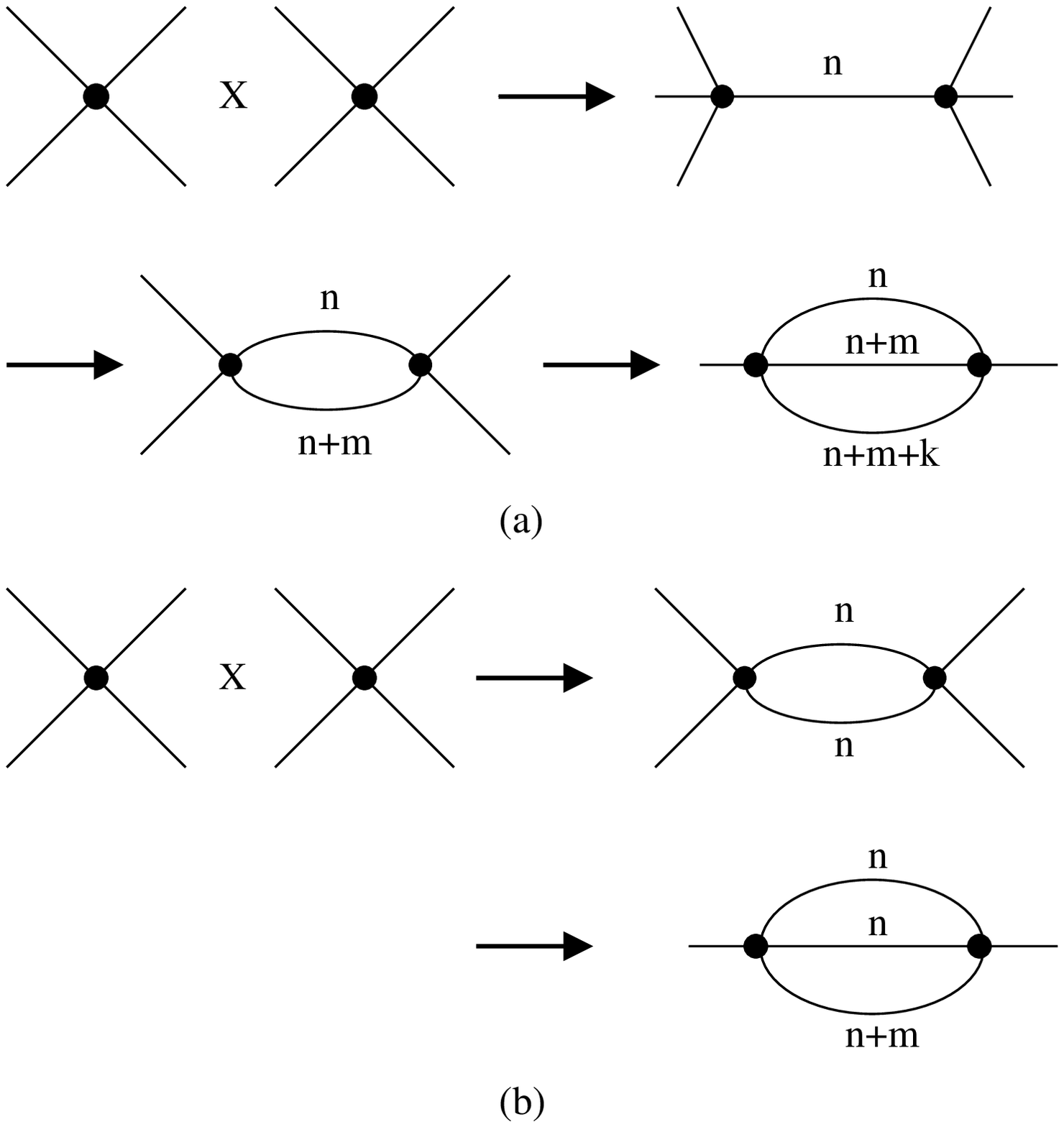}
\caption{
Diagrammatic representation of the two types of cascades of contractions
in the second order over interaction.
(a): 3-step cascade; contractions are done at the steps $n, n+m, n+m+k$.
(b): 2-step cascade; contractions are done at the steps $n$ and $n+m$.}
\end{figure*}

\begin{figure*}
\includegraphics[width=12cm]{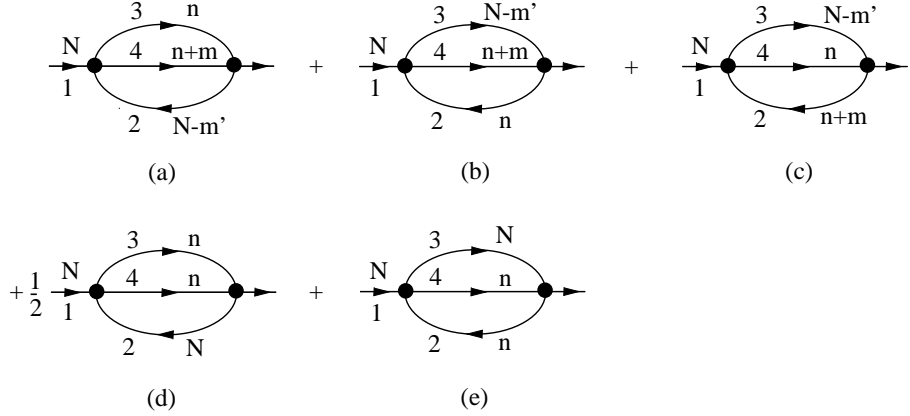}
\caption{
Two-loop kinematically-distinct self-energy diagrams generated after $N$
steps of the Wilson-Kadanoff procedure. Graphs ($a-c$) are generated by
the 3-step cascade (see Fig.~1a), $n <n+m <N-m'$; (d-e) by the 2-step
cascade (see Fig.~1b). Integres shown near the composite variables ($1-4$)
indicate the shell number where the corresponding momenta are restricted to
lie. The signs and combinatorial factors of the graphs stand as they enter
the action's binomial term $\bar\psi(1) \psi(1)$. Implicit sum over all allowed
$n, m (m')$ is assumed.}
\end{figure*}

\begin{figure*}
\includegraphics[width=6cm]{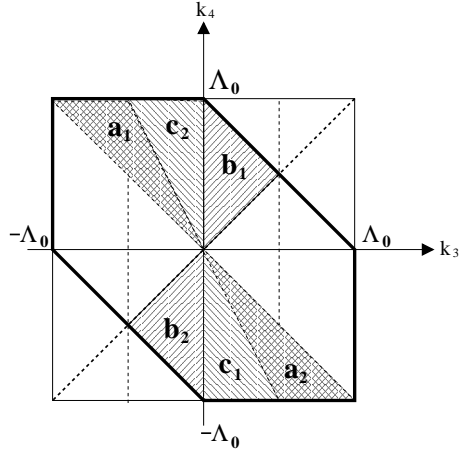}
\caption{
Regions of momenta integrations corresponding to the solutions from Table I
in the limit $\Lambda \to 0$.
The regions for the primed solutions (not given explicitly in Table I)
are obtained by reflection over the main diagonal (bold dashed line),
so the whole set of 12 solutions fill completely the restricted phase space
(\ref{Srestr}), i.e., the hexagon shown by the solid bold line.
The vertical thin dashed lines are guides for an eye. }
\end{figure*}

\begin{figure*}
\includegraphics[width=13cm]{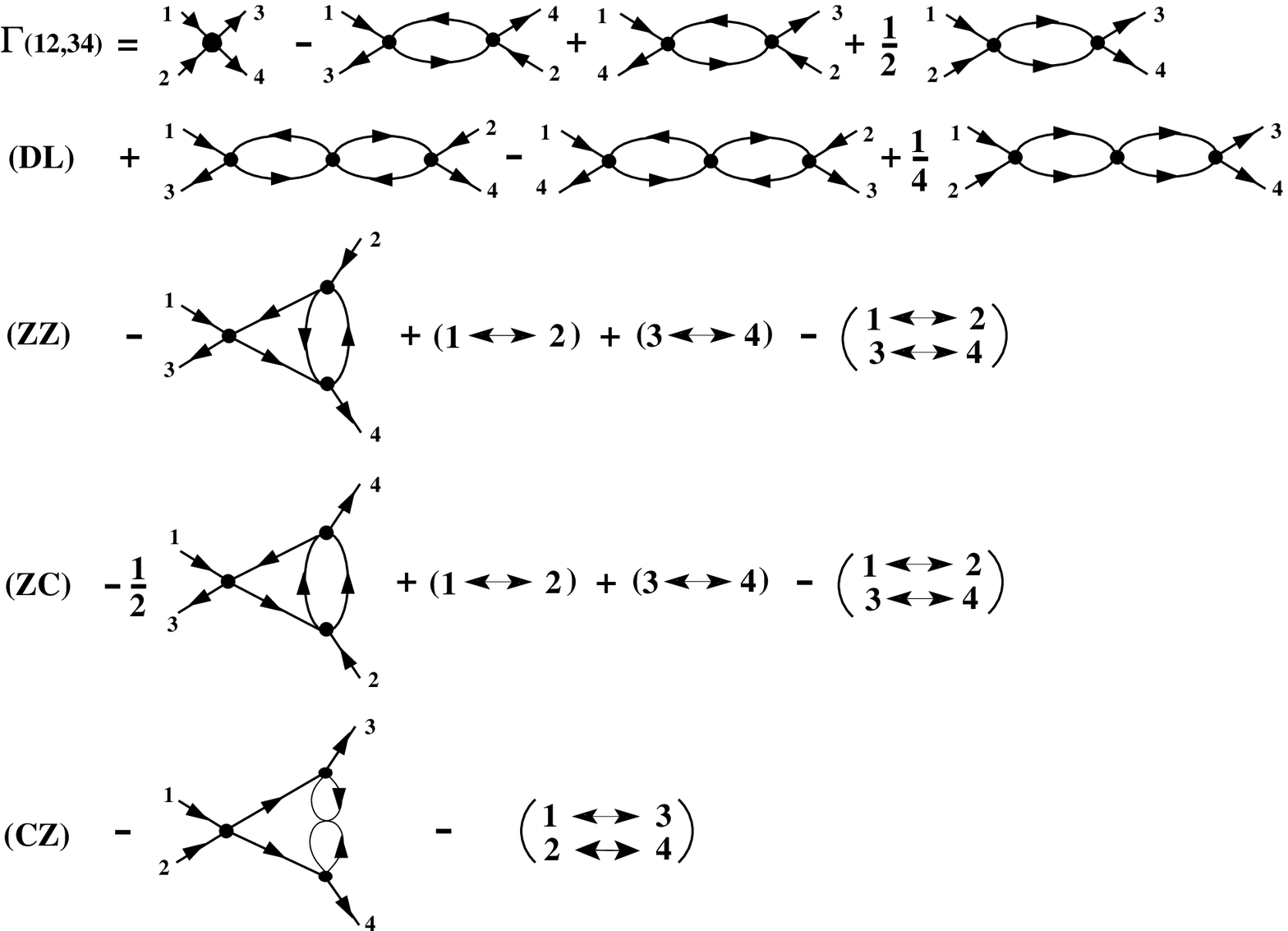}
\caption{
Two-loop perturbative expansion for the four-point vertex.
The tadpole self-energy corrections are absorbed by the Green's
function renormalization.
The vertex variables indicated by numbers comprise the complete
set: chiral and spin indices, Matsubara frequencies, and momenta.
For convenience the two-loop graphs are separated in four groups (DL, ZZ,
ZC, CZ) discussed in the text. Note that in the graph CZ we depicted
a twisted particle-hole loop, not a double loop.}
\end{figure*}

\begin{figure*}
\includegraphics[width=11cm]{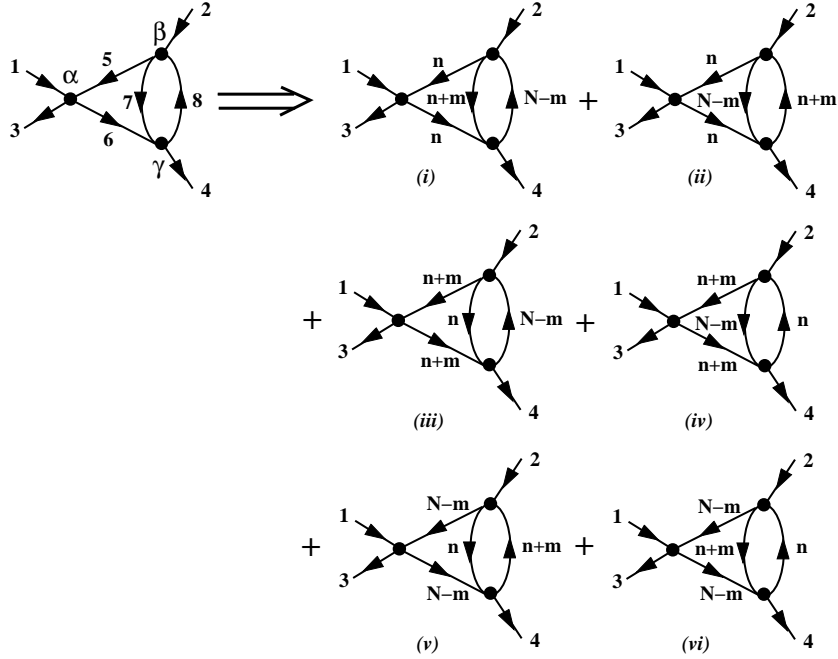}
\caption{
3-step cascades of contactions generating two-loop corrections (topology
of the ZZ, ZC, CZ graphs) to $\Gamma(12,34)$ in the limit (\ref{GamLim}).
As an example, six distinct cascades corresponding to the kinematics
and the external legs configuration of the first graph from the group
ZZ of Fig.~4, are shown. For the sake of space, the vertex labels
and notations for the integrations dummy variables discussed in the text
(the same for all depicted graphs), are explicitly indicated in the top left
one only.
The integers in the graphs \textit{(i-vi)} show the shell numbers where
momenta $(k_5,..,k_8)$ lie. Note that as in the case of the self-energy
corrections (cf. Figs.~1-2), the 3-step cascades come from higher
many-body interactions, themselves generated along the RG flow.
In the graph \textit{(i)}, taken as an example, the first contraction
(step \textit{n}) leads to a four-particle interaction, the second step
\textit{n+m} to a three-particle one, and the final third step \textit{N-m}
yields the two-particle interaction correction \textit{(i)}.}
\end{figure*}
\end{document}